\documentstyle[12pt]{article}
\newcommand{\sptwo}{1.4}

\newcommand{\doublespace}{\edef\baselinestretch{\sptwo}\Large\normalsize}

\begin{document}
\doublespace

$~$
\vspace{55pt}

\begin{center}
{\bf A LOWER BOUND ON NEUTRINO MASS}$^{*\dagger}$\\
Ephraim Fischbach \\
Physics Department, Purdue University, West Lafayette, IN  47907-1396 USA\\
\end{center}
\vspace{3cm}

\begin{center}
{\bf ABSTRACT}
\end{center}

\begin{quote}
The exchange of massless neutrinos between heavy fermions
(e.g. $e,p,n$) gives rise to a long-range
2-body force.  It is shown that the analogous many-body
force can lead to an unphysically large energy density
in white dwarfs and neutron stars.  To reduce the energy
density to a physically acceptable value, the neutrino
must have a {\it minimum mass}, which is approximately
$0.4~eV/c^2$.
Some recent questions relating to the derivation of
this bound are also discussed.

\end{quote}
\vspace{200pt}

\noindent
\underline{$~~~~~~~~~~~~~~~~~~~~~~~~~~~~~~~~~~~~~~~~~~~~~~~~~~~~~~~~~~$}

\noindent
$^*$Work supported in part by the U.S. Department of Energy.

\noindent
$^\dagger$Invited talk at the XXXI Recontres de Moriond, 20-27 January, 1996.

\noindent
To be published in the conference proceedings.

\pagebreak

Two recent papers have re-examined the question of whether
the forces arising from the exchange of $\nu\bar{\nu}$ pairs can be
detected experimentally [1,2].  In the present paper we summarize
the main results in Refs. [1,2], and discuss some more recent
work aimed at clarifying the lower bound on the neutrino mass
derived in Ref.[2].

We begin by considering the self-energy $W$ of a neutron star in
the presence of neutrino-exchange forces, which can be
evaluated by using the following formula due to Schwinger [3]:
$$  W = \frac{i}{2\pi} Tr \left\{ \int_{-\infty}^\infty dE~~\ell n
    [1 + \frac{G_F}{\sqrt{2}} a_n N_\mu \gamma_\mu (1+\gamma_5)
    S_F^{(0)} (E)]
     \right\}.
   \eqno{(1)}
$$  

\noindent
Here $G_F$ is the Fermi constant, $a_n = -1/2$ is the weak $\nu-n$
coupling constant, $N_\mu$ is the vector current of neutrons,
and $S_F^{(0)}$ is the free neutrino propagator.  As we discuss
below, it may be possible to evaluate $W$ for some choices of
$N_\mu$ without expanding $\ell n[1+...]$ in powers of $G_F$,
although for purposes of deriving a bound on the neutrino mass
it is useful to carry out such an expansion.
The contribution proportional to $G_F^k$ arises from a Feynman
diagram in which $k$ neutrons attach to a neutrino loop, as
shown in Fig. 4 of Ref. [1].  Since $G_F$ and $R$ are the only dimensional
variables in the problem, it follows that in order $G_F^k$
the self-energy of a cluster of $k$ particles spread out through
a spherical volume of radius $R$ is proportional to
$G_F^k/R^{2k+1}$.  Since one can form 
{\tiny $ \left(\begin{array}{c}N\\k\end{array}\right)$}
$k$-body clusters from $N$ neutrons, where 
{\tiny $\left(\begin{array}{c}N\\k\end{array}\right)$}$= N!/[k!(N-k)!]$
is the binomial coefficient, it follows that the $k$-body
self-energy $W^{(k)}$ is of order

$$  W^{(k)} \sim \frac{G_F^k}{R^{2k+1}} 
    \left(\begin{array}{c}N\\k\end{array}\right) \cong 
    \frac{1}{k!}\frac{1}{R} \left(\frac{G_FN}{R^2}\right)^k,
    \eqno{(2)}
$$
where we have written 
{\tiny $\left(\begin{array}{c}N\\k\end{array}\right)$}$\cong N^k/k!$
for $k<<N$.  For a typical neutron star $G_FN/R^2 = O(10^{13})$,
and hence it follows from Eq.(2) that higher-order 
many-body interactions make increasingly {\it larger} contributions
to $W^{(k)}$.  Eventually one is led to a paradoxical
situation in which $W = \sum_k W^{(k)}$ exceeds the
known mass $M$ of the neutron star.  If no other mechanism
exists to suppress $W$, then one is led to the 
conclusion that neutrinos must have a minimum mass $m$.
With $m\not=0$ the $\nu\bar{\nu}$-exchange force
``saturates", just as the strong interaction force does,
and for an appropriate value of $m$ the mass-energy of
a neutron star arising from neutrino-exchange would be
reduced to a physically acceptable value.  This minimum
mass (for any species $\nu_e, \nu_\mu,$ or $\nu_\tau$) is

$$  m \stackrel{>}{\sim} \frac{2}{3e^3} \frac{G_F}{\sqrt{2}}
    |a_n|\rho = 0.4~eV,
   \eqno{(3)}
$$
where $\rho$ is the number density of neutrons in
a neutron star, and $\ell n e = 1$.

The detailed calculations which suggest the possibility of
such a bound are presented in Ref. [2].  
We briefly review here a number of possible
questions
that have been raised.  a) One may ask whether perturbation
theory is even valid in the presence of effects as large
as those found in Ref. [2].
This is
addressed in Ref. [2], but an alternative way of viewing
the present calculation is to start with $m\not= 0$.  For
an appropriate value of $m$, the self-energy $W$ would be
acceptably small, and there would be no question concerning
the validity of perturbation theory.  As one attempts to
pass to the $m=0$ limit, $W/M$ eventually exceeds unity,
and this implies that $m$ cannot be smaller
than a certain critical value, namely that given by Eq.(3).
Using this approach one can sidestep various problems in
perturbation theory, and still arrive at the bound in Eq.(3).

b)  Another set of issues relates to the possibility of
calculating $W$ from Eq.(1) without first carrying out a
perturbation expansion.  This can be done, for example, in
the case of an infinite continuous medium with constant mass
density $\rho_m$.  One can anticipate via the following
heuristic argument that in such a system the effects of
neutrino-exchange will be small.
For an infinite system the physically relevant quantity
is the energy density which has dimensions $\mu^4$, where
$\mu$ is a mass scale.  The only available dimensional
quantities are $G_F$ and $\rho_m$ which appear in the
combination $G_F \rho_m \sim \mu^2$.  It follows that
for an infinite medium we expect to find
$$  \mbox {energy~density}~ \sim (G_F\rho_m)^2.
    \eqno{(4)}
$$
This means that the only contribution in the infinite-medium
case is from the 2-body potential, and this conclusion is supported
by detailed calculations.  By contrast, for a neutron star
of radius $R$ with constant number density $\rho$, one can
form the dimensionless quantity appearing in Eq.(2),
$$  G_FN/R^2 = (4\pi/3)G_F\rho R.
   \eqno{(5)}
$$
Since the product $G_F\rho R$ 
is dimensionless it
can appear raised to
any power in the expression for $W$, and this is supported by
both Eq.(2) and the detailed calculation in Ref. [2].
We can infer from this discussion that large cancellations
must take place as one passes from the results for a finite
neutron star to those for an infinite medium.
Moreover, there are some ambiguities in how the infinite-medium
limit is taken, as we show elsewhere.  This discussion
suggests that calculations of neutrino-exchange effects in
an infinite medium may not be directly relevant for a neutron
star of finite radius.

c)  It has been noted [4] that the same combinatoric factors
which enhance the many-body contribution to the self-energy $W$,
also enhance the many-body contribution to the production of
physical $\nu\bar\nu$ pairs.  If these pairs are trapped in
the neutron star, then their presence could serve to Pauli-block
the exchange of the virtual $\nu\bar\nu$ pairs which give rise
to the unphysically large value of $W$.  If this were true,
then we would no longer be forced to the conclusion that
neutrinos must have a minimum mass.  However, there are a
number of problems with the preceding scenario:  i) For
$m=0$ the analogs of the many-body diagrams considered in
Refs. [1,2] would produce both low-energy and high-energy
$\nu\bar\nu$ pairs at an unphysically large rate.
The neutrino star, rather than trapping the neutrinos,
would be destroyed as a result of the large forward
scattering cross section.  ii) Even for low-energy neutrinos
or antineutrinos, the dominant many-body interaction may be
{\it repulsive} (depending on the value of $N$), and
hence the neutron star may {\it expel} both $\nu$ and $\bar\nu$.
(Note that only $k=even$ contributions are non-zero for a
spherically symmetric neutron star, and these produce the
same effects for $\nu$ and $\bar\nu$.)  More generally,
for $m=0$ problems would arise in both the self-energy $W$
and the $\nu\bar\nu$ production rate.  To understand how
these relate to each other and to other processes occurring
in a neutron star (or white dwarf) will require more detailed
calculations.  However, it seems unlikely that 
the mechanism proposed in Ref.[4] can avoid the implications
of Refs. [1,2] that neutrinos must be massive.

I wish to collectively thank my many colleagues for
helpful discussions.

\begin{center}
{\bf REFERENCES}
\end{center}

\begin{enumerate}
\item E. Fischbach, D.E. Krause, C. Talmadge, and D. Tadi\'c,
Phys. Rev. {\bf D52}, 5417 (1995).
\item E. Fischbach, Ann. Phys. (NY) {\bf 247}, xxx (1996).
\item J. Schwinger, Phys. Rev. {\bf 94}, 1362 (1954);
J.B. Hartle, Phys. Rev. {\bf D1}, 394 (1970).
\item A. Yu. Smirnov and F. Vissani, preprint, IC/96/67.
\end{enumerate}

\vspace{20pt}

\begin{center}
{\bf R\'esum\'e}
\end{center}

\begin{quote}
L'\'echange des neutrinos sans masse entre des fermions
lourds produit une force macroscopique entre deux corps.  Nous
montrons que la force analogue entre plusieurs corps m\`ene
a une grande densit\'e de l'\'energie dons les \'etoiles
des neutrons.  Pour r\'eduire la densit\'e de l'\'energie
\`a un valeur acceptable, le neutrino doit avoir une
masse minimum qui est environ de 0.4 $eV/c^2$.
\end{quote}

\end{document}